\documentclass[aps,preprint,amsmath,amssymb,groupedaddress,preprint]{revtex4-2}
\usepackage{amsmath}
\usepackage{graphicx}
\usepackage{color}
\usepackage{color}
\usepackage{graphicx}
\usepackage{natbib}
\usepackage{epsfig}
\usepackage{setspace}
\usepackage{amsmath,amssymb}
\usepackage{verbatim}
\usepackage{bibentry}
\usepackage{xcolor}
\usepackage{bm}
\usepackage{lipsum}

\begin{document}
	\title{Charge and Spin Thermoelectric Transport in Benzene-Based  Molecular Nano-Junctions: A Quantum Many-Body Study}
	
	\author{Parbati Senapati}
	\affiliation{Department of Physics,
		Indian Institute of Technology Patna, Bihta, Bihar, 801106, India}
	\author{Prakash Parida}\email{pparida@iitp.ac.in}
	\affiliation{Department of Physics,
		Indian Institute of Technology Patna, Bihta, Bihar, 801106, India}
	\date{\today}
	
	\begin{abstract}
		{
Within the Coulomb blockade regime, our study delves into the charge, spin, and thermoelectric transport characteristics in a benzene-based molecular nano-junction using Pauli master equation and linear response theory. The charge- and spin-transport studies show strong negative differential conductance features in the current-voltage ($I-V$) characteristics for the ortho and meta connections of electrodes on either side.   Contrarily, the para-connection displays Coulomb staircase behavior.   Exploring spin current behavior in the presence of spin-polarized electrodes or external Zeeman field, we establish a methodology that facilitates precise control over the specific spin flow. 
Various charge and spin thermoelectric transport coefficients have been studied with varying chemical potentials. We focus on spin-polarized- conductance, the Seebeck coefficient, and the figure of merit. By adjusting electrode polarization or employing an external magnetic field, we achieve an impressive peak value for the spin thermoelectric figure of merit, approximately 4.10. This outcome underscores the strategic value of harnessing both spin-polarized electrodes and external magnetic fields within the domain of spin caloritronics.}
\end{abstract}
\maketitle
\section{Introduction}
Molecular electronics is a branch of science that studies the electronic and thermal transport properties of circuits which use individual molecules as basic building blocks. Due to its potential applications in nanoscale electronic devices like transistors, rectifiers, sensors, and switches \cite{perrin2015single,van2010charge,ghosh2021small,martinez2019thermal}, this field of study has garnered a lot of interest from both the research community and industry level. For example, the possibility of investigating electronic and thermal conduction at the smallest  scales, where the usual size of molecules is between 1 and 10 nm, could result in a higher packing density of devices, which would have benefits for cost, efficiency, and power consumption.
The intermolecular interactions that could be used in nanoscale self-assembly technology, possibly leading to low-cost manufacturing, are another appealing feature.
Additionally, chemists have access to a wide variety of molecular structures, which allows them to logically and deliberately alter the properties of molecules.
Molecular junctions are also ideal systems for investigating the fundamental concepts underlying the mechanisms of charge transfer followed by heat (energy) transfer. 
\par
In recent years, researchers have been working on developing high-performance thermoelectric materials and equipment that can recycle waste heat into electricity \cite{wang2005oscillatory,segal2005thermoelectric,lunde2006interaction,markussen2009electron,reddy2007thermoelectricity,chen2015thermoelectric}.  The benefits of thermoelectric (TE) devices, such as their compact size, lack of mechanical moving parts, durability, and ability to produce electricity at low heat gradients, have sparked a lot of interest. Many past attempts have been made to increase thermoelectric performances through experimentation and theoretical work \cite{dresselhaus2007new,zheng2012enhanced,wang2013thermoelectric,liu2010enhancement,boukai2008silicon,wierzbicki2010electric,kuo2011theory,wysokinski2012thermoelectric,dong2015thermospin} using quantum dots and molecules in the Coulomb blockade \cite{beenakker1992theory,turek2002cotunneling,kubala2006quantum} and Kondo regimes \cite{yoshida2009thermoelectric,sakano2007thermopower,dong2002effect,krawiec2006thermoelectric}.
The thermoelectric performance is crucial for describing a thermoelectric device that is important to sustain and limits TE device applicability \cite{wilson2012experimental,kubala2008violation,terasaki1997large}. 
The thermoelectric figure of merit (ZT) determines the performance of a thermoelectric device, $ZT=S^2G_VT/K_t$, where T is the temperature, S is the Seebeck coefficient (thermopower), $G_V$ is the electrical conductance, and $K_t$ is the total thermal conductance given by $K_t=(k_{e}+k_{ph})$, $k_{e}(k_{ph})$ is the electron (phonon) contribution to $K_t$.  Since $ZT$ can be increased by increasing the power factor ($S^2G_V$) or lowering the thermal conductivity, a high-performance thermoelectric material should have both high thermopower and electrical conductivity and low thermal conductivity. However, having a high $ZT$ in natural materials appears challenging. 
Several factors impede the rise of $ZT$. First, the Wiedemann-Franz law \cite{snyder2011complex} is obeyed in typical solids, which means that an increase in electric conductivity leads to an increase in thermal conductivity. Second, the Mott relation \cite{cutler1969observation} states that an increase in electrical conductivity will likely decrease the Seebeck coefficient. Bulk materials in general show low $ZT$. The advancement of nanotechnology has made it possible to enhance the $ZT$ by reducing the dimensional complexity of bulk structures, and it increases the power factor \cite{murphy2008optimal,nozaki2010engineering,hicks1993effect,wang2003spin,mukerjee2007doping}.
Engineering the bulk structure at the nanoscale can induce different aspects regarding the electronic density of states and the transmission coefficient, which help the electrons to pass through a device. Moreover, The Wiedemann-Franz law and other classical conclusions, such as the Mott relation, may not hold because of the quantum phenomena that have evolved in nanostructure materials. The development of the nano-thermoelectric device has sparked interest in this area \cite{wang2013enhancement,karamitaheri2012engineering,o2005electronic}. 

Furthermore, the thermoelectric properties of nanostructure materials can be controlled by varying the gate voltage. As a result, it paves the way for discovering valuable thermoelectric devices on a new and vast scale. In 1993, Hicks and Dresselhaus were the first to exploit low-dimensional structure materials to attain high $ZT$ \cite{hicks1993thermoelectric}. Many theoretical models for thermoelectric transfer via quantum point contacts, quantum dots, and other strongly correlated nanostructures have been proposed \cite{esposito2010efficiency,muralidharan2012performance}. 
Tagani et al. investigate the thermopower of double quantum dot systems that are weakly coupled to metal electrodes using a density matrix approach and observe that an increase in Coulomb repulsion enhances the $ZT$ by lowering the bipolar effect. Additionally, they analyzed how interdot tunnelling affects the figure of merit \cite{tagani2012thermoelectric}. Wierzbicki et al. theoretically investigated the thermal transport via a double quantum dot in the linear response domain. They found that interference effects dramatically increased thermopower and thermoelectric efficiency \cite{wierzbicki2011influence}. Natalya demonstrates how coulomb interactions cause considerable augmentation of the $S$ and $ZT$ in numerous quantum dots in a serial configuration \cite{zimbovskaya2022large}.
\par
Additionally, the exploration and control of spin-polarized current through nanostructures has also attracted a lot of theoretical \cite{busl2010spin,ghosh2015spin} and experimental \cite{engel2004controlling,ding2021cornerstone} attention in recent years due to its significance for our understanding of basic quantum physics and its numerous applications, including spintronics, nanoelectronics, and quantum computation \cite{pulizzi2012spintronics,grover1997quantum}. Researchers are investigating the prospect of using electron spin rather than its charge as the basis of new electronic devices known as ``spintronics'' in an effort to increase computational power and speed. The study of the interaction between spin effects and heat transfer has garnered a lot of attention as spintronics has advanced. It is demonstrated that the relative magnetic configurations of ferromagnetic electrodes affect the thermoelectric effects. A spin-Seebeck effect analogue to the charge-Seebeck effect was recently proposed \cite{cornaglia2012tunable,ramos2017spin}. The experimental technique used for studying spin transport includes scanning probe spin thermoelectric microscopy (SPSTM), which operates by utilizing an STM with a magnetic tip and detects electron spin orientations by sensing the local density of states as the tip nears a sample surface. Applying a thermal gradient induces a spin-dependent Seebeck effect, generating a spin-related thermoelectric voltage. This technique maps these signals with high spatial resolution, investigating spin transport properties at the nanoscale\cite{aradhya2013single}. An experimental measurement of the spin voltage produced by a temperature gradient in a ferromagnetic $Ni_{81}Fe_{19}$ film has also been performed \cite{uchida2008observation}. The observed spin thermopower in this bulk sample, however, is so negligible that it might be dominated by the associated charge thermopower, which is several orders of magnitude more. After that, Dubi and DiVentra \cite{dubi2009thermospin} investigated a quantum dot in contact with two ferromagnetic electrodes that were held at various temperatures and discovered that the spin thermopower can be as great as the charge thermopower and even exceed it in magnitude. The quantum dot needs to be trapped between two ferromagnetic leads in their setup for it to be subjected to a significant amount of Zeeman splitting, which could be limited in terms of practical applicability.
\par
Not limited to quantum-dot nano-junctions, researchers also have experimentally investigated thermoelectric transport properties of various molecular bridges. Many experimental techniques have been employed to study thermoelectric transport in such molecular junctions. In the scanning tunneling microscope break junction (STMBJ) technique, a molecular monolayer is anchored with a bias voltage applied. The tip approaches the substrate, captures molecules, moves away, measures conductance changes, forming electrode-molecule-electrode junctions and providing statistical conductance data\cite{reddy2007thermoelectricity}. To measure thermopower, a temperature difference is induced between the substrate and tip\cite{rincon2016thermopower}. The mechanically controllable break junction (MCBJ) method, similar to STMBJ, adjusts the gap using a piezoelectric actuator, capturing molecules and integrating a heater for thermopower measurement\cite{reed1997conductance,kaneko2015simultaneous}. In thermal conductance assessment, the scanning thermal probe microscopy (SThM) technique, part of atomic force microscopy (AFM), is widely employed, especially for studying thermal conductance in self-assembled monolayers (SAMs)\cite{janus2017micromachined,yue2012nanoscale}. These studies are aimed at revealing the potential of different molecular systems for applications in thermoelectric devices. Diverse examples include organic molecules, inorganic nanowires, single-molecule junctions, as well as other organic and inorganic structures \cite{o2021review,li2023superior,saha2011multiterminal,soto2023theoretical}. This systematic approach enables researchers to assess and predict the thermoelectric performance of a range of molecular architectures, providing valuable insights into their potential applications in energy conversion and electronic devices.

Many theoretical techniques and formalisms have been used for understanding the transport characteristics of single molecules connected to electrodes. The study of transport at the nanoscale has standardised numerical approaches based on the combination of non-equllibrium Green's-function (NEGF) methods with tight-binding models or density functional theory (DFT) \cite{ren2012thermoelectric,ramezani2021side}. These techniques are suitable for analyzing quantum transport in molecular bridges that are strongly coupled to electrodes. In a recent study, Gallego et al. investigated the enhancement of thermoelectric properties in single organic radical molecules through a combined approach of theoretical analysis and experimental validation. The study showcased simultaneous improvements in both the Seebeck coefficient and electrical conductance within Blatter radicals \cite{hurtado2022thermoelectric}. Several groups also have recently considered the thermoelectric potential of conjugated monocyclic compounds like benzene for discovering interference effects in strong coupling regimes. Soto-Gómez et al. investigated the electrical and thermal transport properties of a single-molecule junction of Catechol using real-space renormalized Green’s functions in a tight-binding approximation. Their findings indicate the potential utility of catechol as both a conductive and thermoelectric molecule, opening avenues for innovative applications in molecular electronic devices\cite{soto2023theoretical}. Zhang et al. explored the thermoelectric properties of single molecular junctions using a two-level model based on electron–phonon interactions and NEGF formalism. They investigated the potential for achieving high-efficiency thermoelectric devices by engineering the energy level splitting in the molecular junction\cite{zhang2021thermoelectric}.

Using Green's function technique, Haidong et al. studied the thermoelectric properties of a benzene molecule strongly connected to two metal leads. They found that the magnetic flux significantly impacts the thermoelectric properties. Additionally, they demonstrate how this theoretical benzene model may be applied to high-efficiency thermoelectric devices and has good thermal engine features \cite{li2017enhancement}. In the linear response regime, Sartipi et al. suggested the thermoelectric transport through a benzene molecule strongly connected with three metallic terminals and explored possible conductance and thermopower coefficients. Furthermore, a significant improvement in the figure of merit is seen in the three-terminal configuration with tunable temperature differences. The results indicate that the third terminal model can effectively increase efficiency at maximum output power compared to the two-terminal model \cite{sartipi2018enhancing}.

\par 
However, alternative approaches are needed to describe transport through a molecule that is weakly coupled to leads. Because of the importance of the Coulomb interaction in these systems, it is usual to use a Pauli master equation (PME) for the reduced density matrix in the Coulomb blockade regime. For instance, Hettler et al. used an electronic structure calculation to create an effective interacting Hamiltonian for the benzene orbitals. They used a rate equation technique to determine the I-V properties of the associated molecular junction \cite{hettler2003current}. Begemann et al. investigated the transport properties of a benzene-based single electron transistor. They analyzed how the interplay between orbital symmetry and Coulomb interaction manifests itself as destructive interference involving orbitally degenerate states, resulting in selective conductance suppression and negative differential conductance when the contacts are switched from para to meta configuration \cite{begemann2008symmetry,darau2009interference}.
\par
As discussed above, all previous theoretical studies on transport properties of different molecular bridges either explored only charge transport using both generalized quantum master equation (QME) and NEGF methods \cite{hettler2003current,begemann2008symmetry,darau2009interference,bergfield2009many}, or charge thermoelectric proporties in strong coupling regime using NEGF method \cite{li2017enhancement,sartipi2018enhancing}. To the best of our knowledge, till date, the thermoelectric transport properties of any kind of molecular bridges in the Coulomb blockade (weak coupling) regime where manybody effect is dominant, has not been explored. Choosing benzene molecule as a simplistic yet pivotal toy model, for the first time, we have explored thermoelectric transport in molecular nano-junctions. The versatility of the benzene molecular junction enables us to explore diverse connections (ortho, meta, and para) with electrodes. This selection has yielded valuable insights into distinctive properties associated with each connection.
\par
 Further, the choice of the weak coupling regime for studying charge and thermoelectric transport properties in molecular junctions is motivated by computational and analytical tractability. In the weak coupling limit, the interaction between the molecular system and the electrodes is assumed to be relatively weak compared to other energy scales in the system. This simplification facilitates the use of perturbative methods and allows for more straightforward mathematical treatment. Also understanding and controlling the staircase behavior in weak coupling is crucial for practical applications of single-electron transistors, enabling advancements in ultra-sensitive sensing, quantum computing, and energy-efficient electronics by precise manipulation of individual electrons.  Moreover, in the strong coupling regime, the dominant effects in transport behavior result from the synergy of electrodes, molecules, and their interface. These components collectively shape the overall behavior. However, in a weak coupling regime, the entirety of transport behaviors predominantly originates from the molecule itself. Changes in electrodes or the interface have comparatively minimal impact in such scenarios. 
\par
In this article, using Pauli master equation (PME) formalism, for the first time, we explore the spin- and thermoelectric transport properties in benzene molecular bridges weakly coupled to electrodes on both sides. This method is also extendable and applicable to other molecular systems to study the transport properties in a weak coupling regime. The paper is structured as follows. In section II, we describe the model Hamiltonian and theoretical formulaton for transport studies in our system. In the next section, we present our numerical results and offer rigorous discussions on it. Our results show many interesting transport characteristics of benzene molecular junction in response to spin-polarized electrodes and Zeeman field. The last section includes the final conclusions. 

\section{Theoretical Formalisms}
\subsection{Model}
\begin{figure}
	\centering
	\includegraphics[width=\columnwidth]{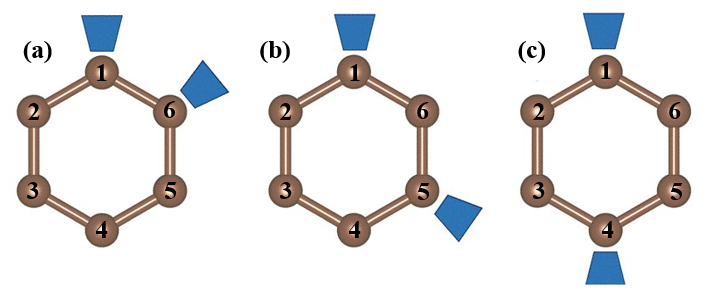}
	\caption{\label{str_diagrm} Representation of the benzene molecule attached to the electrode at (a) ortho (1-6), (b) meta (1-5), (c) para (1-4) connection. Blue blocks represent the electrodes that are weakly coupled to the molecule. }
\end{figure}

We model the benzene molecule using the extended Hubbard model. The benzene molecule is weakly coupled to electrodes at ortho, meta and para connections as shown in Fig. \ref{str_diagrm}. We start by considering only the localized  $\mathrm{P_{z}}$ orbitals (one per carbon atom) in an interacting Hamiltonian of isolated benzene. Considering the nearby electron-electron interactions and hopping between nearest neighbour sites, the most general form of Hamiltonian for benzene is as follows:
	\begin{align}
		H_0 =  \sum_{i=1, \sigma}^N\ \epsilon_{i\sigma} a^{\dag}_{i\sigma}a_{i\sigma} +\sum_{i\sigma } - (t_{i,i+1}a^{\dag}_{i\sigma}a_{i+1,\sigma}+h.c.) \notag\\ 
+U\sum_{i=1}^N(n_{i\uparrow}-\frac{1}{2})(n_{i\downarrow}-\frac{1}{2}) \notag\\
+V\sum_{i=1}^N(n_{i\uparrow}+n_{i\downarrow}-1)(n_{i+1\uparrow}+n_{i+1\downarrow}-1)
     	\label{exhub}
	\end{align} 

\noindent
Where $\epsilon_{i\sigma}$=$\epsilon_{i}-\hat{\sigma}2\mu_{B}B$ is the on-site energy for various atomic sites whose degeneracies are lifted by an external magnetic field B (Zeeman splitting), $\mu_{B}$ is the Bohr magneton, $\hat{\sigma}$=+(-) for $\sigma$=$\uparrow$($\downarrow$), $\epsilon_{i}$ is set to zero for all sites, $t_{i,i+1}$ is the hopping intensity between nearest neighbour sites ($i$ to $i+1$), $U$ is the Hubbard interaction term, and $V$ denotes nearest neighbour coulomb repulsion. $a_{i \sigma}^\dagger$ ($ a_{i \sigma}$) is the creation (annihilation) operator of spin $\sigma$ in the $i^{th}$ ($i=1$ to $6$) site and $n_{i\sigma}$=$a_{i \sigma}^\dagger$$ a_{i \sigma}$. Here, mechanical oscillations are disregarded, and all atoms are assumed in their equilibrium positions. The parameter for carbon-carbon hopping terms is assumed to be 2.5 eV for benzene. According to information from the literature, the Hubbard electron-electron interaction for the electron at the carbon sites and the nearest neighbour coulomb repulsions is assumed to be 10 eV and 6 eV respectively\cite{begemann2008symmetry}.
\subsection{Method}
\subsubsection{Pauli master equation}
For the description of the benzene molecule weakly coupled to electrodes, we utilize the total Hamiltonian $H = H_0+H_{el}+H_T$, where $H_0$ comprising the interacting Hamiltonian for isolated benzene (described in Eq. \ref{exhub}), the Hamiltonian for the electrodes as $H_{el}$ = $\sum_{\alpha k\sigma}\epsilon_{k\sigma}c^{\dag}_{\alpha k\sigma}c_{\alpha k\sigma}$, where $\epsilon_{k\sigma}$ denotes the energy of the electron with wave vector $k$, spin $\sigma$ in the electrode $\alpha$=L/R, $c^{\dag}_{\alpha k\sigma} (c_{\alpha k\sigma}$) are creation (annihilation) operator, and the Hamiltonian $H_T$ = $\sum_{\alpha k\sigma}(b_{\alpha k\sigma}a^{\dag}_{\alpha\sigma}c_{\alpha k\sigma}+b^{\star}_{\alpha k\sigma}c^{\dag}_{\alpha k\sigma} a_{\alpha\sigma}$) representing the coupling between electrodes and molecule. Because of the weak coupling to the electrodes, we can assume that the potential drop is all concentrated at the electrode-molecule interface and has no effect on the molecule itself. The preferred method for addressing dynamics in weak coupling is the Liouville equation approach. The quantum Liouville equation for the reduced density matrix is given by $ \frac{\partial \rho}{\partial t}=\frac{i}{\hbar}[\rho,H]+\mathcal{C}\{\rho\}$, where $\rho$ is the reduced density operator and $\mathcal{C}\{\rho\}$ describes the dissipative part of the dynamics after eliminating electrode memory effect by Markovian approximation (temporal coarse-graining)\cite{gebauer2004current}.

 The reduced dynamics is often approximated by a generalized QME, encompassing both diagonal and off-diagonal elements of the density matrix to account for coherences between different charge states. However, in a weak coupling regime, the validity of the PME\cite{fischetti1998theory,fischetti1999master} is constrained to considering only the diagonal elements of the density matrix in the eigenbasis of the Hamiltonian while neglecting the off-diagonal terms. The PME describes the time evolution of the probabilities of occupation of the eigenstates of the Hamiltonian (diagonal elements of the density matrix)\cite{dubi2009thermospin,hettler2003current}. The standard approach is to take into account the steady-state solution , which reduces the problem to an algebraic system of linear equations, the solution to which is the population of the molecule, or the kernel of the rate matrix. In non-stationary scenarios, the application of the PME would lead to a violation of current continuity\cite{gebauer2004current,frensley1990boundary,novakovic2011quantum}. Moreover, the justification for using the PME is specifically limited to very small devices \cite{gebauer2004kinetic,gebauer2004current}. Given the molecular scale of our system, employing the PME is not only appropriate but also valid in the weak-coupling limit i.e., $\Gamma_0 << k_BT$, (where $\Gamma_0$ is the coupling strength of electrodes taken to be $1~meV$). The weak coupling implies that the symmetry of the molecule, as well as the structure of the Hamiltonian  (Eq. (\ref{exhub})), will remain unchanged.

 PME can be comprehended as an outcome of first-order time-dependent perturbation theory (Fermi's golden rule) or as a solution derived from a many-body Schrödinger equation \cite{bruus2004many}.  The significant advantage of utilizing PME lies in their ability to model extremely complex systems with relative simplicity. These systems would otherwise demand significantly more computational effort or alternative approaches. Moreover, employing PME enables the observation of numerous remarkable phenomena exhibited by such complex systems. We use the exact diagonalization (ED) approach to diagonalize $H_0$ (6-sites Hamiltonian with $4^6$, i.e., 4096 basis) within the PME formalism, while the electrodes are considered as electronic reservoirs and are described by their Fermi distributions. Diagonalization of $H_0$ yields many-body eigenstates  $|s>$ and related eigen energies $E_s$. Electron transfer in molecules and electrodes is exclusively considered within the rate matrix $W$, the elements of which are the rates of transition between many body eigenstates of the reduced system. The PME (taking diagonal element of density matrix ($\rho_{ss}=P_s$)) are as follows\cite{schaller2014open}:
\begin{eqnarray}
	\frac{dP_s}{dt}=\sum_{s^\prime}(W_{{s^\prime} \rightarrow s}P_{s^\prime}-W_{{s} 
		\rightarrow {s^\prime}} P_s).
\end{eqnarray}
\noindent
Where $W_{s^\prime \rightarrow s}$ ($W_{{s}\rightarrow {s^\prime}}$) denotes the rate of transition from the many-body Fock state $s^\prime$ to $s$ ($s$ to $s^\prime$), which differs by one electron and $P_s (P_{s^\prime})$ is the probability that the system will be in the many-body state $s (s^\prime)$. The transition rates are as follows:
\begin{eqnarray}
	W_{{s^\prime}\rightarrow{s}}^{L+}=\Gamma_{L\sigma} f_L(E_s-E_{s^\prime}) \sum_{\sigma} 
	|<s|a^\dag_{1\sigma}|s^\prime>|^2 
\end{eqnarray}
\begin{eqnarray}
	W_{{s^\prime}\rightarrow{s}}^{R+}=\Gamma_{R\sigma} f_R(E_s-E_{s^\prime}) \sum_{\sigma} 
	|<s|a^\dag_{N\sigma}|s^\prime>|^2 
\end{eqnarray}
\noindent
The corresponding equation for $W_{{s} \rightarrow {s^\prime}}^{L-}$ and
$W_{{s} \rightarrow {s^\prime}}^{R-}$ obtained by replacing $f_{L,R}(E_s-E_{s^\prime})$ by $(1-f_{L,R}(E_s-E_{s^\prime}))$,where $f_{L/R}$ is the Fermi function for left/right electrode.  In this case, $+/-$ denotes the creation/destruction of an electron within the molecule due to electron migration from/to the left ($L$)/right ($R$) electrodes. $\Gamma_{L\sigma}$ and $\Gamma_{R\sigma}$ are spin dependent electrode-molecule coupling strength for left and right electrodes. Additionally, we assumed that the creation and annihilation occur only at the sites directly connected to the electrodes. The total transition rate will be calculated by summing four terms ($W_{{s} \rightarrow {s^\prime}}=W_{{s} \rightarrow {s^\prime}}^{L+}+
W_{{s} \rightarrow {s^\prime}}^{R+}+W_{{s} \rightarrow {s^\prime}}^{L-}+
W_{{s} \rightarrow {s^\prime}}^{R-}$).
One can figure out the population of many-body states by solving the steady state PME, $\frac{dP_s}{dt}=\sum_{s^\prime}(W_{{s^\prime} \rightarrow s}P_{s^\prime}-W_{{s}\rightarrow {s^\prime}} P_s) = 0$.  This equation is in the form of a homogeneous linear system (AB=0) that cannot be solved. So, we have used $\mathrm{\sum_s{P_s}=1}$ to eliminate one row/column of the matrix, which helps to reformulate the eigenvector problem into an inhomogeneous linear system (AB=X). This inhomogeneous linear sets of equations can be solved by using linear algebraic methods.
\subsubsection{Charge, spin and heat Currents} 
At a steady state, the current coming from the molecule to one electrode is exactly cancelled by the current flowing from the other into the molecule. However, if only the current between the molecule and one of the electrodes is considered, the expression does not vanish, and we may depict the actual electric(charge) current flowing through the system by, 
\begin{eqnarray}
	I_{\alpha} =\frac{e}{\hbar} \sum_{s,s^\prime}(W_{{s^\prime} \rightarrow s}^{\alpha+}
	P_{s^\prime}-W_{{s} \rightarrow {s^\prime}}^{\alpha-} P_s)
\end{eqnarray}
Similarly, the energy/heat current is also computed as;
\begin{eqnarray}
	Q_{\alpha} =\frac{1}{\hbar}(E_s-E_{s^{'}}) \sum_{s,s^\prime}(W_{{s^\prime} \rightarrow s}^{\alpha+}
	P_{s^\prime}-W_{{s} \rightarrow {s^\prime}}^{\alpha-} P_s)
\end{eqnarray}
where $\alpha=L/R$.
Additionally, the current flowing from the electrode is the sum of current flowing in the spin up and spin down channel. The charge current can be written as, $I_{C}=I_{\alpha\uparrow}+I_{\alpha\downarrow}$. While the spin current flowing from electronic reservoirs (electrodes) are determined by the corresponding spin-polarized charge current i.e.,$I_{S}=I_{\alpha\uparrow}-I_{\alpha\downarrow}$. 

\subsubsection{Thermoelectric formulations}
For thermoelectric investigations, a linear response regime is employed. In this case, we presume that the left lead is slightly hotter than the right one, i.e. $T_L=T_R+\Delta T$ and $\mu_L=\mu_R-e\Delta V$, then the charge and heat current is given by,
\begin{eqnarray}
	I_{\alpha} =G_V \Delta V+G_T \Delta T
\end{eqnarray}
\begin{eqnarray}
	Q_{\alpha} =M \Delta V+K \Delta T
\end{eqnarray}
Where $G_V$ and $G_T$ are electrical conductance and thermal coefficient, respectively. Then we obtain $G_V$ and $G_T$ by putting $I_L=\frac{1}{2}(I_L-I_R)$ and expanding the Fermi-Dirac distribution function as $f_L(x) = f_R(x) - \frac{(x-\mu)}{T} f^{'}(x) \Delta T+ e \Delta Vf^{'}(x)$. where $f^\prime  (x)=\frac{\partial f(x)}{\partial x}$ .
\begin{align}
	G_V =\frac{e^2}{2\hbar}\sum_{s,s^\prime}\sum_{\sigma}\Gamma_{L\sigma} f^\prime  (E_s-E_{s^\prime})|<s|a^\dag_{1\sigma}|s^\prime>|^2 \notag\\ (P_s+P_{s^\prime})
\end{align}
\begin{align}
	G_T =\frac{-e}{2\hbar}\sum_{s,s^\prime}\sum_{\sigma}\Gamma_{L\sigma} f^\prime  (E_s-E_{s^\prime})\frac{((E_s-E_{s^\prime})-\mu)}{T} \notag\\ 
	|<s|a^\dag_{1\sigma}|s^\prime>|^2 (P_s+P_{s^\prime})
\end{align} 
Under the condition of charge current $I_\alpha=0$, thermopower can be calculated as $S=\frac{G_T}{G_V}$. The magnitude and sign of the thermopower are influenced by asymmetry in the distribution of electrons near the chemical potential. Using the Onsager relation, electron thermal conductance can be written as \cite{zianni2008theory},
\begin{eqnarray}
	k_e=\frac{-Q_\alpha}{\Delta T}|_{I_ \alpha=0}\nonumber \\
=K-S^{2}G_{V}T
\end{eqnarray}
Where,
\begin{align}
	K =\frac{1}{2\hbar}\sum_{s,s^\prime}\sum_{\sigma}\Gamma_{L\sigma} f^\prime  (E_s-E_{s^\prime}) \frac{((E_s-E_{s^\prime})-\mu)^2}{T} \notag \\
	|<s|a^\dag_{1\sigma}|s^\prime>|^2 (P_s+P_{s^\prime}) 
\end{align}
Then the figure of merit can be written as, $ZT=\frac{S^{2}G_{V}T}{k_e+k_{ph}}$ .The phonon thermal conductance can be expressed as $k_{ph}=3k_0$ ($k_0=(\pi^2 k_B^2 T)/3h$  is the quantum thermal conductance).

\section {Numerical Results and Discussion}
Our work introduces a novel dimension by pioneering the investigation of charge and spin thermoelectric transport within benzene molecular junctions using the PME method. In our quest to comprehend these transport mechanisms, we delve into the analysis of charge and spin current transport within the junction. While many previous research has delved into this realm, our approach distinguishes itself through the incorporation of uniquely formulated equations and parameters. This strengthens the robustness of our research endeavor. 

\subsection{Charge and spin Transport}
In our investigation, we have meticulously examined both charge and spin transport properties within the benzene molecule, encompassing ortho, meta, and para connections. This molecule has been weakly coupled to electrodes positioned on either side. To ensure the molecule remained in its six-electron ground state with no bias, we skilfully tuned the chemical potential of the electrodes. This adjustment of the chemical potential prevented any net charge transfer between the molecule and the electrodes, maintaining the desired neutral charge state and facilitating a zero-bias environment. Employing the PME approach, we have conducted a comparative analysis of the current-voltage ($I-V$) characteristics across three distinct connections of the benzene molecule. These connections represent the ortho, meta, and para configurations. Notably, our analysis is solely focused on the neutral (six-electron) or anionic (seven electron) charge states of the benzene molecule. The nature of current-voltage characteristics in nanoscale systems is intricately influenced by a multitude of factors. Among these, the electronic structure of the constituent molecule stands out as a pivotal determinant. This electronic arrangement governs the availability and movement of charge carriers, thereby shaping the overall transport behavior. Equally vital is the underlying physics that composes charge and energy flow within the system. Quantum mechanical effects, electronic interactions, and coupling between the molecule and the electrodes collectively contribute to the observed current-voltage traits. Furthermore, the profile of the potential drop experienced across the molecule as it spans the gap between the electrodes plays a critical role. This electrostatic potential landscape directly affects the energy levels and pathways available to charge carriers, thereby influencing the flow of current and the resulting voltage response.
\par
In summary, the intricate interplay of these factors—the electronic structure of the molecule, the underlying physical mechanisms, and the electrostatic potential distribution—exerts a profound influence on the observed current-voltage characteristics in nanoscale systems. Understanding and controlling these elements are paramount in designing and tailoring the functionality of such systems for a diverse range of applications.
\par
To elucidate the origins of all observed current-voltage ($I-V$) characteristics, our analysis delves into several crucial factors. These encompass a detailed examination of the charge density distribution across all molecular sites from which we gain insights into how charge carriers are distributed throughout the molecule, the eigenstates corresponding to the 6-electron ground state ($\phi_{6}^{gs}$), and the 7-electron ground state with both spin-up ($\phi_{7\uparrow}^{gs}$) and spin-down ($\phi_{7\downarrow}^{gs}$) configurations which provides a comprehensive view of the energy landscape and the accessible electronic transitions and lastly the occupation probabilities of all relevant electronic states, a comprehensive grasp of how charge carriers populate the available energy levels within the system. A schematic depiction of the energy landscape associated with the transition from $6e^{-}\rightarrow7e^{-}$ states in benzene is presented in Fig. S1, available in the supporting information (SI). This transition represents the lowest energy excitations in low-bias scenarios.

\subsubsection{Case of Normal electrode ($\Gamma_{L\sigma}$=$\Gamma_{R\sigma}$=$\Gamma_{0}$, B=0)}

\begin{figure*}
	\centering
	\includegraphics[width=1.0\textwidth]{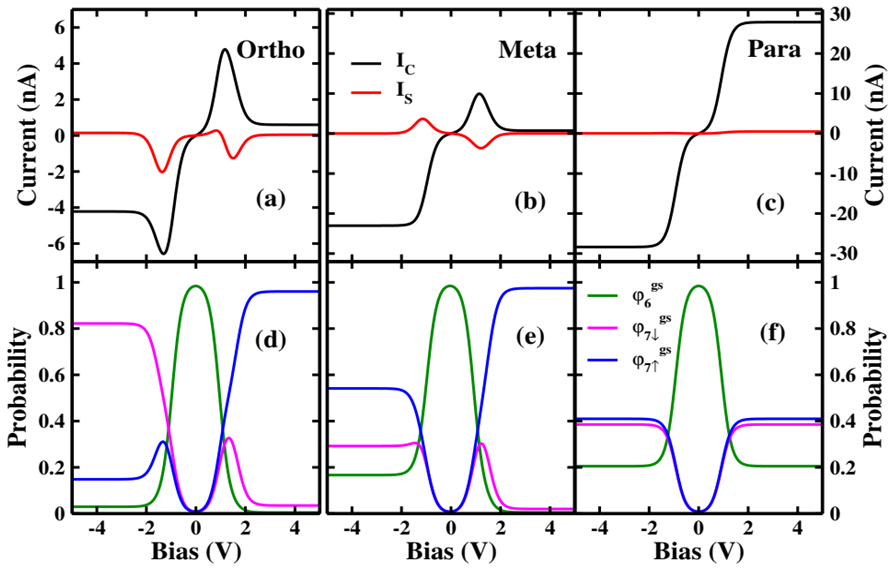}
	\caption{\label{NE_cur} (a), (b) and (c) represents the charge ($I_C$) and spin ($I_S$) current of ortho, meta and para connection of benzene respectively, (d), (e) and (f) represents the respective probabilities of occupied many-body electronic states for the normal electrode connections. The probabilities of occupied electronic state are represented by a green line for $\phi_{6}^{gs}$ (singlet), a magenta line for $\phi_{7\downarrow}^{gs}$ (doublet) and a blue line for $\phi_{7\uparrow}^{gs}$ (doublet).}
\end{figure*}

\par
Fig. \ref{NE_cur} depicts the charge ($I_C$) and spin ($I_S$) current of benzene molecular junction as a function of bias voltage with three different configurations (i.e., ortho, meta, and para) of normal electrode (NE) connection. With the unique geometric arrangements and electronic properties associated with each configuration, we anticipate the emergence of intriguing and distinctive patterns in the way current responds to the applied bias voltage. It is noticed that the ortho $(1-6)$ and meta $(1-5)$ connections of benzene exhibit remarkable negative differential conductance (NDC)-like features in $I_C$. The distinct nature of this behavior is visually depicted in Fig. \ref{NE_cur} (a) and \ref{NE_cur} (b), where an increase in bias leads to the abrupt decline of current. 
The emergence of NDC can be directly attributed to the distinct coupling geometry present in the ortho and meta connections of the benzene molecule, as thoroughly investigated in our study. In these specific molecular arrangements, the intricate interplay between the molecular structure and the applied bias voltage gives rise to a remarkable phenomenon. As the bias voltage increases, this unique geometry enforces a nonconducting state that becomes increasingly prominent beyond a specific threshold voltage. Essentially, this particular coupling geometry controls the flow of charge through the molecular junction. This discovery underscores the connection between molecular architecture and electronic behavior, emphasizing how specific configurations can generate intriguing and impactful transport characteristics. 
\par
Understanding the pivotal role played by molecular geometry in shaping electronic transport not only advances our theoretical insights but also opens doors to designing tailored molecular devices with precise and desirable electronic functionalities-a promising frontier in the realm of molecular electronics. 
Similar NDC characteristics for benzene have already been reported in the literature, including radiative relaxation to a blocking state \cite{hettler2003current} and an interference-induced blocking state \cite{darau2009interference,begemann2008symmetry}. NDC has also been demonstrated experimentally for molecular junctions with a wide range of molecules, including polyporphyrine oligomers, nitroamine-functionalized benzene, and azobenzene\cite{chen1999large,kuang2018negative,choi2006conformational}. Unlike the NDC-like features observed in the ortho and meta connections, the current-voltage ($I-V$) characteristics of the para ($1-4$) connection of the benzene molecule displays a distinctive staircase-like behavior, as illustrated in Fig. \ref{NE_cur} (c).  This staircase-like response reinforces the relation between the molecular structure, electronic states, and bias voltage.
\par
The analysis of occupation probabilities of various many-body states within the ortho connection of benzene, as depicted in Fig. \ref{NE_cur} (d), provides valuable insights into transport behavior. At bias voltages ranging from $-0.2$ to $0.2$ V, the occupation probabilities predominantly favour the 6-electron ground state ($\phi_{6}^{gs}$), which is a singlet, while other states remain nearly unpopulated. This bias range is insufficient to induce an extra electron into the molecule, thus maintaining a stable neutral charge state. As a consequence, no current flows through the molecular bridge, and the system operates within the Coulomb blockade regime.
As the bias voltage increases, both in the negative and positive regimes, the probability of occupying the 7-electron ground states with spin up ($\phi_{7\uparrow}^{gs}$) and spin down ($\phi_{7\downarrow}^{gs}$) rises. We note here that 7-electron ground state is a 4-fold degenerate state: $\phi_{7\uparrow}^{gs}$ (doublet) and $\phi_{7\downarrow}^{gs}$ (doublet). This increased probability renders these states energetically accessible for electron transitions from the 6-electron state to the 7-electron state, facilitating a current flow.
Remarkably, in the bias range of $1.18$ to $2.12$ V, a distinct change occurs in the occupation probabilities. While the probability of $\phi_{7\downarrow}^{gs}$ state diminishes, that of $\phi_{7\uparrow}^{gs}$ state continues to rise. This variation in probabilities correlates with the emergence of NDC in the current-voltage characteristics. Importantly, in this region, the probability of $\phi_{7\uparrow}^{gs}$ state dominates over $\phi_{6}^{gs}$ and $\phi_{7\downarrow}^{gs}$ states, signifying that $\phi_{7\uparrow}^{gs}$ state predominantly contributes to the NDC behavior observed in the positive bias regime of the ortho connection of benzene.
Conversely, the origin of NDC in the negative bias regime ($-1.29$ to $-2.10$ V) is distinct. Here, the probability of $\phi_{7\downarrow}^{gs}$ state predominates, leading to the occurrence of NDC. The spin current, as shown in Fig. \ref{NE_cur} (a), demonstrates a negative value in the positive bias regime due to the dominance of the current contributed by the down electron ($I_{\downarrow}$) over the current from the up electron ($I_{\uparrow}$). Conversely, in the negative bias region, $I_{\uparrow}$ significantly outweighs $I_{\downarrow}$, resulting in a negative spin current for that regime.

\par
Analyzing the occupation probability distribution within the context of the meta-connection of benzene (illustrated in Fig. \ref{NE_cur} (e)) yields insightful observations about its transport behavior. At very low bias ranges, the $\phi_{6}^{gs}$ state becomes predominantly occupied. As the bias voltage is applied in both the positive and negative regimes, the probabilities associated with $\phi_{7\uparrow}^{gs}$ and $\phi_{7\downarrow}^{gs}$ states rise, rendering them energetically accessible for electron transitions and subsequently enabling the flow of current.
However, a notable transition becomes evident in the positive bias regime within the range of $1.13$ to $2.23$ V. As bias increases within this window, the probability associated with the $\phi_{7\downarrow}^{gs}$ state diminishes, while that of the $\phi_{7\uparrow}^{gs}$ state experiences a rapid increase. This pronounced shift in probabilities is intimately tied to the emergence of NDC in $I_C$.
The behavior of the spin current exhibits distinct characteristics in the positive and negative bias regimes, primarily due to the fact that the current associated with the down electron ($I_{\downarrow}$) outweighs the current attributed to the up electron ($I_{\uparrow}$) in both these regimes.

\begin{figure}
	\centering
	\includegraphics[width=\columnwidth]{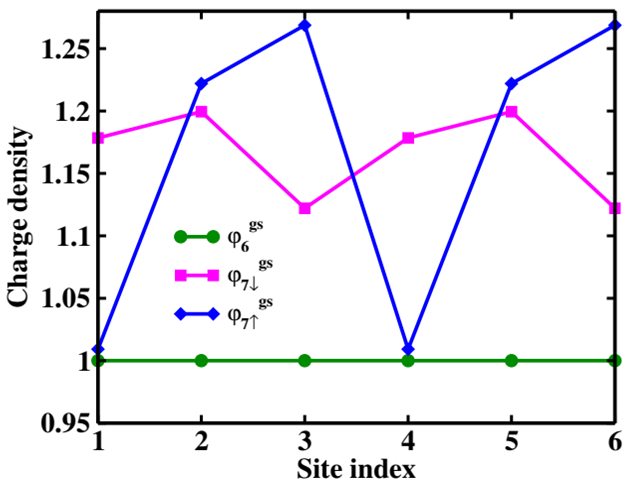}
	\caption{\label{chrg_den} The charge density distribution of 6-electron ground state ($\phi_{6}^{gs}$), spin up and spin down of  7-electron ground state ($\phi_{7\uparrow}^{gs}$ and $\phi_{7\downarrow}^{gs}$) over the sites of benzene.}
\end{figure}
\par
Fig. \ref{NE_cur} (f) provides a visual representation of the probability analysis for the para connection of benzene. At bias voltages ranging from $-0.4$ to $0.4$ V, the 6-electron ground states ($\phi_{6}^{gs}$) are fully occupied, placing the system within the Coulomb blockade region. Consequently, during this bias range, the system remains nonconductive, yielding no current flow.
However, as the bias voltage increases beyond this range, an intriguing transition occurs. Both the 7-electron ground states with spin up ($\phi_{7\uparrow}^{gs}$) and spin down ($\phi_{7\downarrow}^{gs}$) become increasingly active in the transport process. As the probabilities associated with these states rise and the occupancy of $\phi_{6}^{gs}$ decreases, the conditions become favorable for the transition of an electron from a 6-electron state to a 7-electron state. This transition facilitates the onset of current flow through the system.
Furthermore, with a continued increase in bias voltage, the transport process is influenced by the involvement of higher electronic states, leading to a steep rise in current. Notably, in this case, the observed spin current is minimal. This is attributed to the nearly equal contributions of the currents associated with spin up ($I_{\uparrow}$) and spin down ($I_{\downarrow}$), resulting in a balanced overall spin current behavior.
By delving into the occupation probability analysis and spin current behaviors within the ortho, meta, and para-connection of benzene, we gain a comprehensive understanding of the interplay between electronic states, bias voltage variations, and transport phenomena. This knowledge contributes to a deeper comprehension of the phenomenon of NDC as well as the staircase behaviors, enriching our understanding of molecular electronic transport dynamics.

\par

The distribution of charge density across all sites of the benzene molecule has been thoroughly investigated for both the $\phi_{6}^{gs}$ state and, the $\phi_{7\uparrow}^{gs}$ and $\phi_{7\downarrow}^{gs}$ states, as depicted in Fig. \ref{chrg_den}. This analysis provides crucial insights into the population of molecular sites and their propensity to facilitate the creation and annihilation of electrons, ultimately driving charge transport. The significant influence of the onsite Hubbard electron-electron interaction leads to a single electron occupying each site within the $\phi_{6}^{gs}$ state.
To comprehensively grasp the implications of this charge density distribution and gain a deeper understanding of the nature of the $I-V$ characteristics of the molecules, let us delve into an interpretation. For simplicity, we will assume that the electrochemical potential at the left electrode diminishes with a bias of $\mu-V/2$ and increases with a bias of $\mu+V/2$.
As the bias increases, it can be seen that transition $6e^{-}\leftrightarrow7e^{-}$ occurs which implies that an electron is created at the right electrode resulting in a current flow. Again, when the bias increases, the transition towards a more excited state may occur. However, in our study, we are only concerned with low-bias characteristics.
\par
In the $\phi_{7\downarrow}^{gs}$ state, the electrons remain almost equally distributed over all the atomic sites. Consequently, an electron jumps from the $\phi_{6}^{gs}$ to the $\phi_{7\downarrow}^{gs}$ and has been created at the right electrode with an increase in bias. As a result, the current increases significantly. But the $\phi_{7\uparrow}^{gs}$ exhibit unequal charge density distribution of electrons in ortho and meta connection which is not favourable for electron transition. Therefore, the current diminishes as bias rises in the positive regime, leading to an NDC in $I-V$ characteristics. Again, in the ortho connection, there is an NDC in a negative bias regime because a little bit of difference has been noticed in the $\phi_{7\downarrow}^{gs}$ state. In the para connections, $\phi_{7\uparrow}^{gs}$ as well as $\phi_{7\downarrow}^{gs}$ state have almost equally distributed electrons in all atomic sites, giving a step like $I-V$. As a result, the population and depopulation of the anionic states, where the charge distribution patterns are distinctive and important for current flow, produce an exhilarating non-linear behaviour of current-voltage characteristics.
\begin{figure}
	\centering
	\includegraphics[width=\columnwidth]{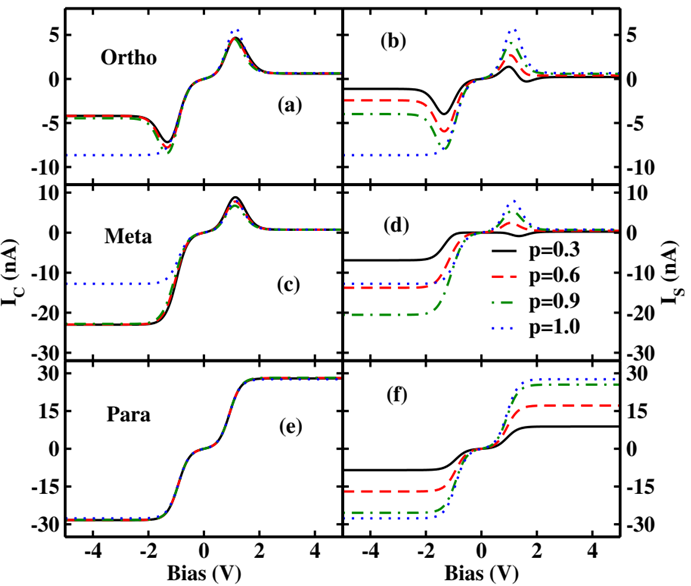}
	\caption{\label{ferro_curr} The charge ($I_C$) and spin ($I_S$) current as a function of the bias voltage in the ortho, meta and para connection of benzene molecule at different spin polarization $p$. }
\end{figure}
\subsubsection{Case of ferromagnetic electrode ($\Gamma_{\alpha\uparrow}$$\neq$$\Gamma_{\alpha\downarrow}$) }

In this section, we delve into the impact of ferromagnetic electrodes (FE) weakly coupled to distinct configurations of the benzene molecule. The presence of ferromagnetism in the electrodes introduces a spin-dependent coupling strength, denoted as $\Gamma_{\alpha\sigma}$, where $\alpha$ represents the electrode and $\sigma$ denotes the spin orientation. This coupling strength is elucidated through a spin-polarization parameter $p$, formulated as $p=\frac{\Gamma_{\alpha\uparrow}-\Gamma_{\alpha\downarrow}}{\Gamma_{\alpha\uparrow}+\Gamma_{\alpha\downarrow}}$ \cite{feng2005spin}.
The spin-dependent nature of these coupling strengths plays a crucial role in the electron transport properties of the molecular junction. This coupling introduces an intriguing avenue to explore spin-related effects and their interplay with molecular configurations and bias voltage. Further analysis will shed light on the interdependence between these factors, offering insights into the intricate dynamics of charge and spin transport in these systems.
\par
In our study, we have exclusively considered the parallel configuration of electrodes, characterized by $\Gamma_{L\uparrow(\downarrow)}=(1\pm p)\Gamma_{0}$ and $\Gamma_{R\uparrow(\downarrow)}=(1\pm p)\Gamma_{0}$. Fig. \ref{ferro_curr} illustrates the charge and spin currents of the ortho, meta, and para connections of the benzene molecule as a function of bias voltage, considering different degrees of polarization.
Fig. \ref{ferro_curr} (a), (c), and (e) present the charge current behavior, revealing that as the polarization $p$ increases, there are subtle or even negligible changes in the peak height/position of the current curves across the three distinct benzene connections. However, as expected and shown in Fig. \ref{ferro_curr} (b), (d), and (f), the behavior of the spin current undergoes a drastic change with increasing polarization. Notably, the spin current displays a rapid ascent as polarization increases. This trend signifies that the absolute value of the spin-up current ($I_{\uparrow}$) consistently surpasses the absolute value of the spin-down current ($I_{\downarrow}$). This phenomenon can be attributed to the accumulation of electrons with a specific spin direction $\sigma$ on the atomic sites of the molecule within specific bias windows.
These distinctive outcomes in the spin current ($I_S$) can also be understood in the context of the values of $\Gamma_{L\sigma}$ and $\Gamma_{R\sigma}$, which play a significant role in shaping the overall transport behavior. Taking all these factors into consideration, it becomes evident that, as polarization increases, the localization of spin-up electrons onto the molecular atomic sites becomes progressively more pronounced compared to spin-down electrons \cite{souza2007quantum}. The interplay of polarization, electrode-molecule coupling, and spin characteristics leads to the fascinating and complex spin-related behaviors observed in the current-voltage characteristics.

\subsubsection{Case of normal electrode with external magnetic field (B) }
\begin{figure}
	\centering
	\includegraphics[width=\columnwidth]{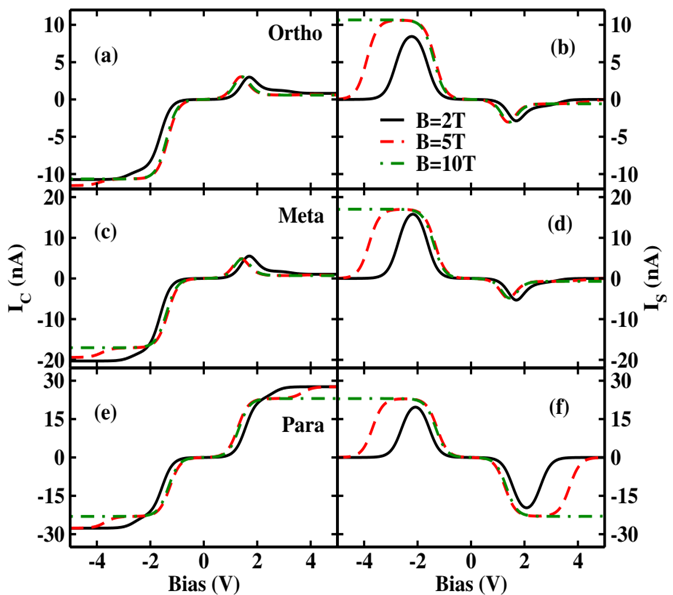}
	\caption{\label{NE_mag_fld} The charge ($I_C$) and spin ($I_C$) current as a function of the bias voltage in the ortho, meta and para connection of benzene molecule at different magnetic field $B$. }
\end{figure}
 In this section, we shift our focus to examine the influence of an external magnetic field on the weakly coupled normal electrodes connected to the molecule. The presence of an external magnetic field leads to the splitting of the molecular energy levels, thereby facilitating the selective transport of pure spin-up ($\uparrow$) or spin-down ($\downarrow$) electrons through the system. This effect can introduce significant spin-related phenomena and behaviors that contribute to the overall transport characteristics in the molecular setup. 
\par
Fig. \ref{NE_mag_fld} displays the charge current ($I_C$) and spin current ($I_S$) as functions of bias voltage (V) in the ortho, meta, and para connections of the benzene molecule under varying external magnetic fields. In Fig. \ref{NE_mag_fld} (a), (c), and (e), the charge current responses exhibit subtle shifts as the external magnetic field strength is increased. This phenomenon can be attributed to the energy level adjustments prompted by changes in the magnetic field. When the magnetic field strength is changed, the energy levels experience shifts, causing them to move away from the bias window through which the current flows. Consequently, different bias voltage windows might be necessitated to facilitate current flow under different magnetic field conditions. 
In addition, the behavior of specific energy levels within the molecular system is crucial to understand the effects of the external magnetic field. At the $\phi_{6}^{gs}$ energy level, the total spin is zero. Consequently, this energy level remains unaffected by alterations in the magnetic field (charge density plot for the various magnetic fields shown in Fig. S2 in SI).
However, the introduction of a magnetic field leads to the disruption of 4-fold degeneracy within the 7-electron ground state, $\phi_{7}^{gs}$ (Table S1 displays the energy level splitting in the $7e^{-}$ state with variations in the magnetic field has been provided in SI). This disruption is evident in the form of an energy level splitting, resulting in the emergence of two distinct doublet states: $\phi_{7\downarrow}^{gs}$($7e^{-}$ ground state) and $\phi_{7\uparrow}^{1es}$ ($7e^{-}$ 1st excited state). The extent and direction of this energy level splitting depend on the orientation of the external magnetic field.

\par
Fig. \ref{NE_mag_fld} (b), (d), and (f) display the spin current responses as a function of bias voltage for three distinct benzene configurations. Notably, these results underscore the remarkable ability to manipulate the transport of a specific type of electron, either spin $\uparrow$ or spin $\downarrow$, by adjusting the strength of the external magnetic field.
When applying magnetic fields of $B=2T$ and $5T$, a clear trend emerges: the spin-down current ($I_\downarrow$) surpasses the spin-up current ($I_\uparrow$). This observation signifies that a greater number of spin-down electrons become localized at specific atomic sites within the molecule, thus contributing to the dominant current flow of spin $\downarrow$ electrons.
Intriguingly, when a stronger magnetic field, $B=10T$, is introduced, a fascinating outcome occurs. All three benzene configurations exhibit the flow of pure spin-down current ($I_\downarrow$). This distinctive behavior can be comprehended through the concept of energy levels that align with the specific bias window enabling current conduction. Under this strong magnetic field, the eigenstate associated with $\phi_{7\downarrow}^{gs}$ remains within this bias window, whereas the $\phi_{7\uparrow}^{1es}$ state experiences a greater shift, causing it to fall outside this bias window. Consequently, the system exclusively facilitates the flow of spin-down electrons through the molecular junction.
This capability to selectively manipulate and control the flow of a specific spin current holds significant promise, especially in the context of spintronic applications. The introduction of an external magnetic field offers a versatile tool to tailor the transport characteristics of molecular systems, which can potentially revolutionize the development of advanced spintronic devices.
Indeed, the findings strongly suggest that utilizing an external magnetic field to modulate the behavior of a weakly coupled molecule to a normal electrode can offer distinct advantages compared to the alternative of switching to a spin-polarized electrode (more flexibility in adjusting the strength and orientation of the magnetic field).

\subsection{Charge thermoelectric transport}

We then investigate the thermoelectric transport properties of the meta connected benzene with varying temperature and the chemical potential. Because the position of $\mu$ is a crucial ingredient in understanding different transport across molecules. It is especially interesting since changing the $\mu$ by doping, adding a side group, a gate field, or using another contact material can optimise molecular characteristics. In a small system, we have also verified and previously published research \cite{beenakker1992theory} findings that changing the $\mu$ of the electrodes and connecting a gate voltage in the channel (molecule) produce the same results. Here we investigate all results by varying chemical potential. In the following section, we delve into the thermoelectric properties of the meta connection of benzene. This emphasis is placed due to the fact that the results from the other two connections, namely ortho and para, yield nearly indistinguishable outcomes. Consequently, our focus remains directed towards explaining the fundamental physics underlying these phenomena and exploring the potential utility of benzene molecular junctions in upcoming fields such as electronics, optoelectronics, and thermoelectrics.
\par

\begin{figure*}
	\centering
	\includegraphics[width=1.0\textwidth]{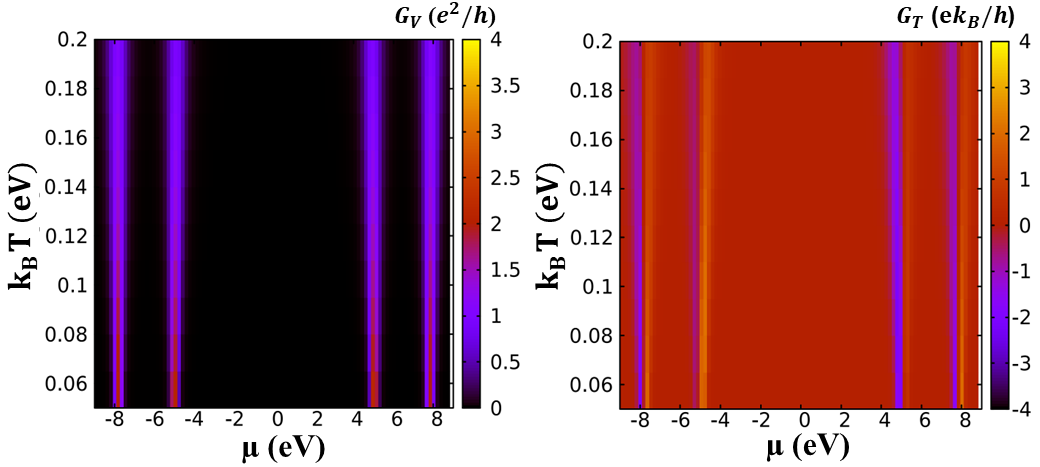}
	\caption{\label{gv_gt} Electrical conductance ($G_V$) (left panel) and thermal coefficient ($G_T$) (right panel) as a function of chemical potential ($\mu$) at different temperatures.}
\end{figure*}

The electrical conductance of benzene molecule as a function of $\mu$ at different temperature is shown in the left panel of Fig. \ref{gv_gt}. As all calculations were performed with a zero bias, the transition between the lowest energy levels (i.e., only the ground state with $N$ and $(N\pm1)$ particles with significant probabilities of occupied different many-body states) is important for conductance. The electrical conductance ($G_V$) peaks occur when the energy required for an electronic transition, specifically from the $Ne^-$ state to the $(N+1)$ $e^-$ state, aligns with the chemical potential of the electrode. The $G_V$ shows four sharp peaks (two peaks in negative and two peaks in positive $\mu$ (found also in Fig. S3, in SI)) corresponding to resonance in meta-connected benzene. The first peak appears at $\mu$ $\approx$ -$7.84$ $eV$. As the $\mu$ increases, the second peak appears arround $-5.05$ $eV$. A positive change in $\mu$  resulting in the emergence of the third peak in conductance at $\approx$ $5.05$ $eV$ (which is the transition energy between $6e^-$ to $7e^-$ states,  $\delta\varepsilon$= $E_7-E_6$ = $5.05$ $eV$).
At a very low-temperature limit, as the Fermi-Dirac distribution function approaches a step function, its derivative ($f^\prime(\delta\varepsilon)$) exhibits a sharp peak (Dirac delta type) when the energy level resonant with chemical potential ($E\approx\mu$). In our case, $f^\prime(\delta\varepsilon)$ is expected to show a sharp peak at $\mu$=$\delta\varepsilon=5.05 ~eV$. When this sharp peak aligns with the energy levels associated with electronic transitions in a system, it can lead to peak in electrical conductance. In Fig. S5 ( found in SI), $G_V$ shows a peak at $\mu$= $5.05$ $eV$,  where $f^\prime(\delta\varepsilon)$ also exhibits a sharp peak. Moreover, when $f^\prime(\delta\varepsilon)$ is at its maximum, we noticed that the probabilies of $6e^-$ ($P_6$) and probabilies of $7e^-$ ($P_7$) states are equal i.e. $P_6$=$P_7$=0.5. At $\mu$= $5.05$ $eV$, the only transition possible from 6$\rightarrow$7 indicating that only $6e^-$ and $7e^-$ states are populated. Solving $\frac{dP_s}{dt}=0$, further confirms that only $P_6$ and $P_7$ are dominated. According to the detailed balance condition, $P_6$ and $P_7$ are equal probable and emerges peaks in conductance. The final peak appears when the transition from $7e^-$ to $8e^-$ configuration occurs, where the probability of 7- and 8-electron configurations is dominated.  
\par
Indeed, in the proximity of the resonance points, the populated states mainly involve the $N$ and $N\pm1$ particle configurations. This phenomenon is closely tied to the conductance behavior of benzene, where specific energy transitions become dominant, leading to pronounced conductance peaks. Additionally, the presence of Coulomb blockade regions between these peaks is a noteworthy feature.
An intriguing and notable observation is the symmetry exhibited by the conductance peaks concerning the chemical potential. This symmetry indicates that changes in the chemical potential result in a proportional adjustment of the conductance peaks, maintaining a consistent pattern. This intriguing symmetry emphasizes the relationship between electronic energy levels, chemical potential, and the resulting conductance characteristics. Such symmetry could hold significant implications for designing and optimizing molecular electronic devices based on benzene and similar molecular systems.

Additionally, we observed that the conductance resonance peaks diminish with wider intensity as the temperature rises (due to the influences of Fermi-Dirac distribution) which suggests that the resistance of the system rises. At low temperatures, the Fermi-Dirac distribution results in a sharp occupancy of energy levels up to the chemical potential. When the temperature rises, the broader distribution of energy levels means that a wider range of energy states is populated by electrons. This leads to a situation where the conductance resonance peaks, which were initially sharp at specific energy levels, become less distinct and spread out over a broader energy range.
Indeed, a significant change in the conductance characteristics of a material or system can have a substantial impact on its thermoelectric properties, specifically the thermopower (also known as the Seebeck coefficient) and the figure of merit (ZT). When the thermopower increases and combines with appropriate electrical conductivity and low thermal conductivity, the ZT value can be enhanced. This results in improved thermoelectric efficiency, making the material more suitable for applications like waste heat recovery or power generation. 
\begin{figure*}
	\centering
	\includegraphics[width=1.0\textwidth]{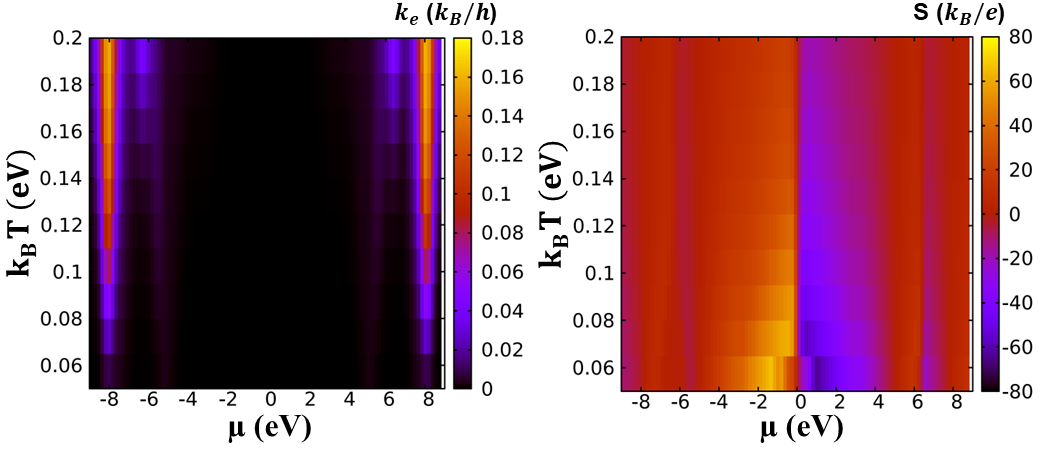}
	\caption{\label{thermo_coeff} Electronic thermal conductance ($k_e$) (left panel) and Seebeck coefficient ($S$) (right panel) as a function of chemical potential ($\mu$) at different temperatures. }
\end{figure*}

The right panel of Fig. \ref{gv_gt} depicts the behaviour of the thermal coefficient as a function of $\mu$ at different temperatures. It has been noted that $G_T$= 0 for some $\mu$ (-7.84,-5.05,0,5.05 and 7.84 eV). First, when $G_V$ attains the peak value, $G_T$ and $S$ are zero because an electron transition occurs at that specific $\mu$ (as discussed above), resulting in an electrical current but no net energy transport. Here the probability of being in the $N$ and $N+1$ electron states will be equal to each other and the transition energy for both electron configurations exactly matches with $\mu$ of the electrode. The temperature gradient does not significantly produce the charge current at resonance energies.  Additionally, in benzene, there is no $G_T$ in the plateau region (where there is no electrical conductance) of $G_V$, indicating that there won't be any current due to temperature gradient if there is no carrier to flow. The $G_T$ changes sign at all resonance points indicating that both type of carriers i.e., electrons and holes are involved in the transport mechanism. The conductance is caused by the transfer of holes when $G_T$ is positive, and when the conductance is caused by the transfer of electrons, $G_T$ yields negative values.  Finally, we also observe that, as the temperature rises, the width of the $G_T$ expands, but the intensity of the $G_T$ decreases due to the temperature dependence on Fermi-Dirac distribution.

\par
Materials with low thermal conductance are used to create a compelling and reliable thermoelectric device. Fig. \ref{thermo_coeff}(left panel) displays the electronic thermal conductance as a function of $\mu$ at different T. 
At low temperatures ($T\approx2K$) \cite{swirkowicz2009thermoelectric}, the peaks of $G_V$ and $k_e$ appear when the transition energy overlaps with the resonant energy; as a result, there is some heat and charge current transfer due to electrons. When resonance approaches, the charge current starts to flow and develop a peak in electric conductance. The electron contribution to heat transfer is then primarily due to electrons resonantly tunneling through the system, and heat conductance approximately follows the charge conductance. However, at high temperatures, the difference between $G_V$ and $k_e$ becomes apparent because as the temperature rises, the thermal distribution of electrons becomes broader. Because of this distribution, tunneling electrons contribute differently to charge and heat conductance due to distinct energy-dependent weighting (energy of a tunneling electron is insignificant for charge current but crucial for heat transfer). Since we calculate all of the results here at high temperatures, $k_e$ behaves differently than $G_V$, and the peaks of thermal conductance are merged to generate some smaller peaks in $k_e$ of benzene, as shown in Fig. \ref{thermo_coeff} (left panel). In the coulomb blockade region, the charge and heat conductances are suppressed because of no carrier flow. As the thermal conductance of benzene is less, it should be more efficient and reliable for future thermoelectric device applications. Moreover, we also observed that, as the temperature rises, the electronic thermal conductance also rises steeply.
\begin{figure}
	\centering
	\includegraphics[width=1.0\textwidth]{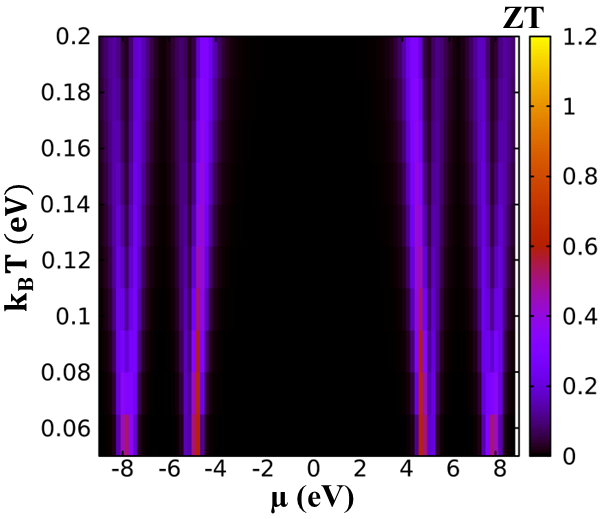}
	\caption{\label{zt} Thermoelectric figure of merit (ZT) as a function of chemical potential ($\mu$) at different temperatures. }
\end{figure}

\par
The $\mu$ and $T$ dependence of the thermopower is shown in the right panel of Fig. \ref{thermo_coeff}. The saw-tooth pattern (2D pattern is shown in Fig. S4 in SI) of the thermopower indicates a change in the electron population of the molecule. The sawtooth pattern observed in the Seebeck coefficient as a function of chemical potential can also be attributed to discrete electronic energy levels within the molecular junction. As the chemical potential varies, electronic states become successively occupied or unoccupied, leading to changes in the charge carrier concentration and, consequently, the Seebeck coefficient. The discrete nature of the energy levels results in abrupt shifts in the Seebeck coefficient, creating a sawtooth pattern. The thermopower sign at different $\mu$ determines the type of dominant charge carriers (hole or electrons). The positive (negative) thermopower suggests that holes (electrons) are responsible for the current. When the $\mu$ crosses the symmetry point ($\mu$=$0$) or resonant quantized levels, the thermopower (or Seebeck coefficient) changes magnitude and sign alternately, exhibiting oscillating behavior with $\mu$. In Fig. \ref{thermo_coeff}, S changes the sign near the symmetry point and reaches a sharp maximum on one side of the symmetry point and a minimum on the other. When $\mu$ deviates further from the symmetry point, S changes sign once more. The Coulomb blockade effect is responsible for the underlying physics and this dependence is consistent with other research insights \cite{beenakker1992theory,turek2002cotunneling,koch2004thermopower,swirkowicz2009thermoelectric}. Therefore, $S$ is zero in these energies (symmetry point and resonant quantized levels). As $\mu$ increases and approaches the resonance, electrons tunnel from the higher temperature electrode to the lower temperature electrode, causing a voltage drop under the condition of diminishing current. As a result of this voltage drop, the thermopower S is positive (S is negative in units of $k_{B}/e$ due to the negative charge of the electrons). At resonance, the induced voltage drop is zero and the thermopower vanishes because electrons can tunnel to the molecule from the colder and hotter electrodes without requiring a temperature gradient. When $\mu$ is less than the resonance, the net electron flow is from right to left and the sign of S changes. The thermopower vanishes and changes sign again in the middle of the Coulomb blockade regime, as previously mentioned. The other resonance has been also explaining in the similar way. In the  symmetry points, electrons and holes have the same weight but opposite signs in the transport mechanism, so the net current is zero. In other words, electrons and holes carry the charge in opposite ways. There is non-zero thermopower at some specific $\mu$, indicating that only one type of dominant charge carrier has participated in transport.
\par 
The probabilities of occupied many-body electronic states near the $\mu$ are crucial factors in determining the thermoelectric properties of a system. The Seebeck coefficient is influenced by the asymmetry in the probabilities of occupying different many-body electronic states with respect to $\mu$. An asymmetric distribution of occupied states can lead to a non-zero or even high Seebeck coefficient. In Fig. S3 (found in SI), we observed that, at $\mu$ = $5.05 ~eV$, $G_V$ attains a peak where $G_T$ and $S$ are zero. This occurs when $\delta\varepsilon = \mu$, and P$_6$ and P$_7$ are equally probable. The condition $\delta\varepsilon = \mu$ indicates resonance, where the energy required for an electronic transition matches the chemical potential. When this alignment occurs, electrons can effectively transition between the $6e^-$ and $7e^-$ states. However, since $P_6$ and $P_7$ are equal, there is no net energy transport. The absence of energy transport can be attributed to the fact that, on average, the energy gained during transitions from $6e^-$ to $7e^-$ is balanced by the energy lost during the reverse transitions from $7e^-$ to $6e^-$. Consequently, while there is electron transport, there is no net energy transfer, leading to $G_T = 0$ and $S = 0$ at $\mu$ = $5.05 ~eV$. The widening peak in the Seebeck coefficient with increasing temperature is attributed to the thermal broadening of energy levels, influenced by the Fermi-Dirac distribution. This thermal broadening effect highlights the intricate interplay between temperature, energy distribution, and the resulting thermoelectric properties in the studied system.    

\par
Finally, we investigate the thermoelectric performance of the molecular junction. As illustrated in Fig. \ref{zt}, the figure of merit is zero at resonances (because $ZT$ is directly dependent on $S$) and coulomb blockade region. Due to the inverse relationship between conductance and thermopower, the figure of merit exhibits fluctuating behaviour. Furthermore, because $S$ and $G_V$ decrease with increasing temperature, $ZT$ decreases with broader peaks shown in Fig. \ref{zt}. The distinctive advantage inherent to molecular systems is their ability to make a negligible phononic contribution to thermal conductivity ($k_{ph}$). This unique characteristic endows these systems with the ability to predominantly influence thermal performance through electronic thermal conductivity ($k_{e}$). $ZT$ has the maximum peak value at low temperatures when $k_e$ is very low. As benzene have lower thermal conductance, it exhibits more enhanced figure of merit which is applicable for thermoelectric performance. It was demonstrated that the activated contribution to $k_{ph}$ can significantly boost thermoelectric performance at relatively low temperatures. Consequently, this study underlines the promising prospects of utilizing benzene molecular junctions to pave the way for groundbreaking advancements in thermoelectric devices. By capitalizing on the enhanced figure of merit and manipulating the interplay between electronic and phononic contributions to thermal conductivity across distinct temperature ranges, these molecular systems hold immense potential for revolutionizing thermoelectric technologies.

\subsection{Spin thermoelectric transport}

In this section, we have investigated the spin thermoelectric transport in our molecular junction  (meta-connection) in a Zeeman-type external magnetic field as well as switching the electrode from normal to ferromagnetic. This intricate configuration possesses the distinctive capability of facilitating pure spin thermoelectric transport phenomena within the molecular junction.
In order to calculate the spin-Seebeck coefficient, we consider a system with both an infinitesimal temperature bias and a spin-voltage bias. The charge and spin currents are described as $I = I_{\uparrow} + I_{\downarrow}$, and $I_{s} = I_{\uparrow} - I_{\downarrow}$ (note: they have the same dimensions). In linear response, the spin current is given by $I_{s}=G_{V}^{s}\Delta V+G_{T}^{s}\Delta T$, where the thermal response coefficient $G_{T}^{s}$ is connected to the fact that a temperature gradient can cause both a spin flow and an energy flow. When we set $I_s$ to zero, we get the spin-Seebeck coefficient $S_s=\frac{G_{T}^{s}}{G_{V}^{s}}$. The spin figure of merit can be defined as  $Z_{s}T=\frac{G_{V}^{s}S_{s}T}{k_{e}^{s}+k_{ph}}$. The spin conductance $G_{V}^{s}$  may be negative; hence the absolute value is used. In the same way that charge transfer is expected to be good, a system with $Z_{s}T > 1$ is expected to be a good heat-to-spin-voltage converter.
\begin{figure}
	\centering
	\includegraphics[width=\columnwidth]{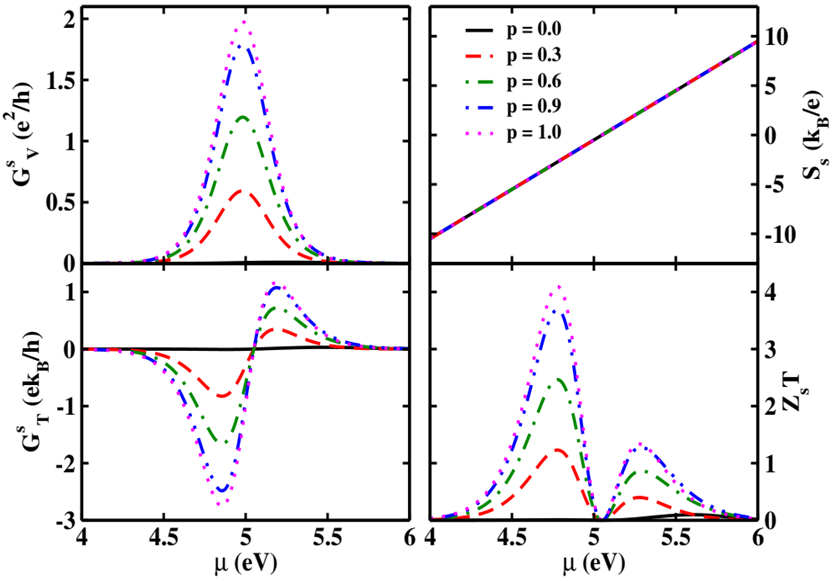}
	\caption{\label{ferro_spin_thermo} (a) Spin conductance $G_{V}^{s}$, (b) spin Seebeck coefficient $S_s$  (c) spin thermal coefficient $G_{T}^{s}$, and (d) spin thermoelectric figure of merit $Z_{s}T$ as a function of chemical potential ($\mu$) at different polarization (p) in meta-connected benzene.}
\end{figure}
In our investigation, an intriguing observation emerged: a spin thermoelectric effect manifested even in the absence of spin-polarized electrodes. Initially, we noticed that the ground state would lead to an equal electron distribution, as evidenced by the charge density plot in Fig. \ref{chrg_den}. Remarkably, we strategically selected a parameterized system that yielded a perceptible spin-thermoelectric effect, all without resorting to alterations in the ferromagnetic electrodes or the application of an external magnetic field. This intriguing outcome strongly hints at the intricate interplay between charge distribution, electronic configuration, and spin-related phenomena. The localized or delocalized nature of electronic states within the connections of benzene might be pivotal in generating such spin-dependent transport behavior, even without explicit spin polarization of electrodes. Our carefully designed parameterized system have the different delocalization of electronic charge state in $\phi_{7\uparrow}^{gs}$ and $\phi_{7\downarrow}^{gs}$ in meta-connected benzene ( Fig. \ref{chrg_den}), showcasing this unexpected spin-thermoelectric effect, underscores the inherent propensity of the system. The absence of external influences, such as magnetic fields, further emphasizes the unique attributes of the system. Additional simulations or experiments with varying system parameters could unveil valuable insights into the underlying mechanisms governing this spin thermoelectric effect. This discovery opens intriguing possibilities, potentially paving the way for innovative energy conversion strategies and novel functionalities by harnessing spin-related phenomena in thermoelectric applications.

Subsequently, we employ ferromagnetic electrodes to delve into the realm of spin thermoelectric transport within benzene molecular junctions. These ferromagnetic electrodes, coupled with the molecule, serve as efficient electron reservoirs exhibiting pronounced spin polarization. Fig. \ref{ferro_spin_thermo} offers a comprehensive depiction of all thermoelectric coefficients, revealing their dependencies on $\mu$ across various polarizations (p). Given our prior analysis, we shift our focus in this section towards examining the involved influence of spin-polarized electrodes on the spin thermoelectric effect. We can see in Fig. \ref{ferro_spin_thermo} (a), (c), and (d) that $G_{V}^{s}$, $G_{T}^{s}$, and $Z_sT$ strongly vary with the spin polarization of electrodes and increase with p, whereas the spin Seebeck coefficient (Fig. \ref{ferro_spin_thermo}(b)) is unaffected by p. The spin-dependent coupling strength is a multiplicative constant. Consequently, as polarization increases, both $G_{V}^{s}$ and $G_{T}^{s}$ experience proportional increase, given their role in governing interactions between particle spins and external influences. The parameter $S_s$ is intricately linked to the $\frac{G_{T}^{s}}{G_{V}^{s}}$ratio. So, when polarization increases, the relationship between $G_{V}^{s}$ and $G_{T}^{s}$ remains consistent. This leads to the observation that alterations in polarization do not impact the value of $S_s$. Moreover, the spin Seebeck coefficient is a material-specific property that is determined by the material's intrinsic properties and the temperature gradient applied. It doesn't depend on the spin polarization of the electrodes in the same way that other thermoelectric coefficients might, because it is more related to the intrinsic properties of the material and its response to temperature gradients rather than the specifics of the electrode configuration.
Furthermore, the sign of $G_{T}^{s}$ and $S_s$ is opposite to the sign of $G_T$ and $S$ as the voltage induced in the channel corresponding to down spin is higher than up spin. Hence, a significant discovery emerges: the strategic integration of a higher spin-polarized electrode offers distinct advantages in the realm of spin caloritronics, leading to a notable rise in the value of $Z_sT$ (maximum value of $Z_sT$ with $p=1$ is 4.09) as the degree of spin polarization (p) increases (shown in Fig. S6 in SI). The maximum value of $Z_sT$ is notably high compared to  to earlier studies on quantum dots i.e., around 0.17 \cite{dubi2009thermospin}, 0.6 \cite{swirkowicz2009thermoelectric}.  
 This underscores the inherent benefits associated with enhancing spin polarization, ultimately amplifying the effectiveness and potential of spin-based caloritronic processes.
\par
\begin{figure}
	\centering
	\includegraphics[width=\columnwidth]{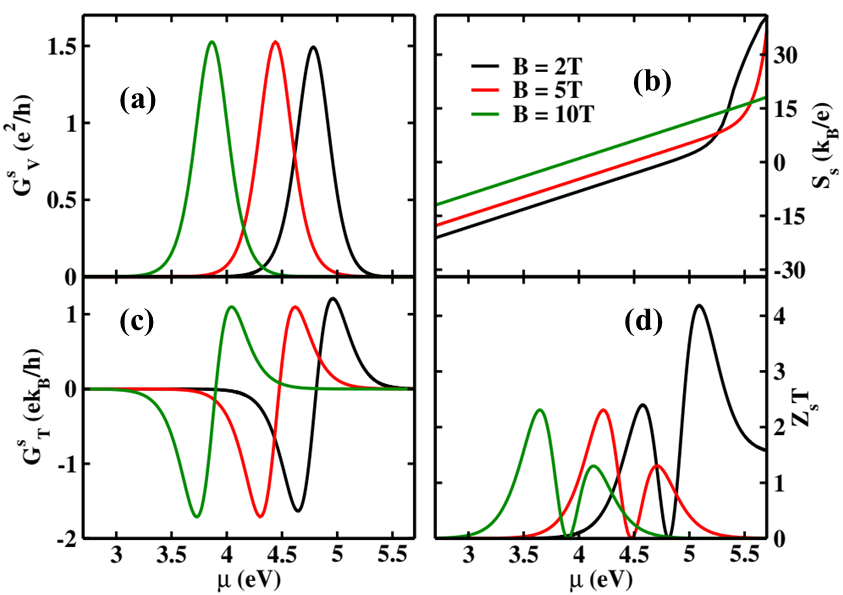}
	\caption{\label{diff_B} (a) Spin conductance $G_{V}^{s}$, (b) spin Seebeck coefficient $S_s$, (c) spin thermal coefficient $G_{T}^{s}$, and (d) spin thermoelectric figure of merit $Z_{s}T$ as a function of chemical potential ($\mu$) at different $B$ in meta-connected benzene.}
\end{figure}
We also explore the influence of an external magnetic field on the spin thermoelectric properties. Fig. \ref{diff_B} depicts the thermoelectric coefficients as a function of $\mu$ at various magnetic field (B). Initially, upon introducing a magnetic field of strength 2T, a noteworthy observation emerged. The resulting maximal value of $Z_sT$ approximated 4.10, which  aligning remarkably well with the $Z_sT$ value witnessed in exclusively pure spin-polarized electrodes ($p=0$). In contrast to spin polarized electrodes, we observed an interesting trends here, i.e.  in Fig. \ref{diff_B}  (a), (c), and (d), we see that the maximum value of $G_{V}^{s}$, $G_{T}^{s}$, and $Z_sT$ has been saturated with increasing magnetic field (also shown in Fig. S7 in SI). The saturation of these parameters suggests that while magnetic fields can certainly influence spin thermoelectric properties, there exists an upper threshold beyond which their impact on transport behavior becomes less pronounced. As the magnetic field increases, the energy levels can become more broadened due to the Zeeman effect. Strong magnetic fields can enhance the filtering of spins with certain orientations, leading to a polarization of the spin current. In our case, when B=10T, the occupancy of the spin-down level increases and tunneling processes involving the spin-down channel become more efficient. This means that electrons with a spin-down orientation find it easier to tunnel through the molecular junction. This is a significant observation, as it indicates a preference for electron transport with a specific spin orientation under the influence of the external magnetic field. However, this effect can saturate as the magnetic field becomes stronger and all available spins of a particular orientation are already aligned and accounted for. This saturation phenomenon serves as a critical factor in comprehending the relation between magnetic fields and the thermoelectric response of molecular systems.\\
One thing we also noticed is that the peak positions shift when the magnetic field changes in comparison to the other cases mentioned above. In the presence of an external magnetic field, the energy levels associated with different spin orientations experience splitting. Consequently, the electronic structure of the molecular system is modified, leading to shifts in the positions of peaks and valleys in transport properties.
 In conclusion, our investigation has illuminated the advantageous role of spin-polarized electrodes as well as the application of an external magnetic field within the realm of spin caloritronics. This discovery holds promise for future innovations in spin-based energy conversion and manipulation technologies. By using spin-polarized electrodes, we achieve efficient spin filtering, allowing only specific spin orientations to pass through the molecular junction. This can lead to a higher degree of spin polarization in the charge carriers, enhancing the potential for exploiting spin-dependent thermoelectric effects. External magnetic fields offer the advantage of direct control over the spin dynamics within the molecular junction. This control can be valuable for investigating the interplay between spin states and thermoelectric properties.

\section{Conclusion and outlook}
Our investigation into charge transport properties within benzene molecular junctions across different connections (ortho, meta, and para) has yielded insightful findings. Notably, ortho and meta connections exhibit prominent negative differential conductance within specific voltage ranges, holding potential for novel single-molecule electronic and thermionic devices. Prior to the implementation of spin-polarized electrode, our study uncovers a noteworthy phenomenon: the influence of spin current even in the absence of an external magnetic field. This intriguing behavior emerges due to our precisely designed parameterized regime, resulting in discernible charge localization and delocalization in the anionic charge states. Our study also explores the effect of ferromagnetic electrodes and external Zeeman field on spin current. Applying a strong external magnetic field to the system enables precise control of specific spin transport (spin up or spin down). This proves more advantageous for achieving pure spin current flow compared to adjusting spin-polarized electrodes, as it offers greater flexibility in tuning the strength and orientation of the magnetic field.
\par 
Further, our exploration extends to investigate the thermoelectric transport properties of the molecular junction in the linear response regime. The observed trends in response to varying chemical potentials showcasing low thermal and high electrical conductance signify a promising avenue for refining thermoelectric capabilities.  Important spin thermoelectric coefficients (i.e., spin conductance, spin Seebeck coefficient and the spin figure of merit) with an external magnetic field as well as spin-polarized electrodes have been studied. By tuning electrode polarization or applying an external magnetic field, we achieve a remarkable maximum spin thermoelectric figure of merit value, approximately 4.10. This value is notably high compared to previous studies involving quantum dots. These results underscore the strategic advantage of utilizing both spin-polarized electrodes and external magnetic fields in the realm of spin caloritronics.
\par
 While single-molecule devices within a weak coupling regime have limitations, it's crucial to explore their behavior, especially in scenarios where electron-electron correlation dominates. In such instances, the weak coupling regime remains applicable despite its limitations. In a recent study, Li et al. highlighted the limitations of the mean-field-based DFT+NEGF approach, attributing its failure to address strong correlation effects in the spin-crossover molecule. They instead utilized a Coulomb blockade model, characterized by weak coupling, which effectively aligned with the experimental results\cite{li2022negative}. Fu et al. investigated the transport behavior of single-molecule Field-Effect Transistors (FETs), providing detailed insights into phenomena such as Coulomb blockade, the Kondo effect, and electron–phonon coupling. The study also explored diverse applications of single-molecule FETs, including the regulation of quantum interference, spin effects, thermoelectric effects, and superconductivity\cite{fu2022recent}.
  Ongoing research in the molecular-scale led to promising applications in various areas such as single-molecule transistors, spintronics devices, thermoelectric nanogenerators, molecular-scale temperature sensors, miniature cooling devices, quantum computing thermal management, and numerous others\cite{perrin2015single,misiorny2009spin,selzer2006single}. In conclusion, our study not only contributes crucial insights into the intricate interplay of charge, spin, and thermal transport at the molecular scale but also paves the way for exploring other molecular junctions for potential thermoelectric applications.

\section*{acknowledgement}
PS acknowledges DST-INSPIRE (IF190005) for financial support. PP thanks DST-SERB for ECRA project (ECR/2017/003305).

\bibliography{thermo-many-body}
\bibliographystyle{rsc}

\end{document}